**Title:** *Scale-Free Exponents of Resting State provide a Biomarker for Typical and Atypical Brain Activity*

Short Title: Scale-Free exponents of Resting State

**Corresponding Author**: SJ Hanson

　　　　　　　　　　　RUBIC

　　　　　　　　　　　Rutgers University

　　　　　　　　　　　197 University Street

　　　　　　　　　　　Newark, NJ 07052

**Telephone:** *973-353-3317*

**Email:** *jose@rubic.rutgers.edu*

word count: 6006

tables: 2

figures: 8

supplemental: 1

**Title:** *Scale-Free Exponents of Resting State are Biomarkers of Neuro-Typical and Atypical Brain Activity*

**Authors:** S.J. Hanson, D. Mastrovito, C. Hanson, J. Ramsey & C. Glymour

**Affiliations:** Rutgers University (RUBIC), Carnegie Mellon University

**Abstract:**

Scale-free networks (SFN) arise from simple growth processes, which can encourage efficient, centralized and fault tolerant communication (*1*).   Recently its been shown that stable network hub structure is governed by a phase transition at exponents (**>2.0)** causing a dramatic change in  network structure  including a  loss of global connectivity,  an increasing minimum dominating node set, and a  shift towards increasing connectivity growth compared to node growth.   Is this SFN shift identifiable in atypical brain activity?  The  Pareto Distribution (P(**D**)~**D**^-β) on the hub Degree (**D**) is a signature of scale-free networks.  During resting-state, we assess Degree exponents across a large range of neurotypical and atypical subjects. We use graph complexity theory to provide a predictive theory of the brain network structure.  *Results.* We show that neurotypical resting-state fMRI brain activity possess scale-free Pareto exponents  **(1.8 se .01)** in a single individual scanned over 66 days as well as in 60 different individuals (**1.8 se .02**).  We  also show that 60 individuals with Autistic Spectrum Disorder, and 60 individuals with Schizophrenia  have significantly higher  (>2.0) scale-free exponents **( 2.4 se .03, 2.3 se .04)**, indicating more fractionated and less controllable dynamics in  the  brain networks revealed in resting state.  Finally we show that the exponent values vary with phenotypic measures of atypical disease severity indicating  that the global topology of the network itself can provide specific diagnostic biomarkers for atypical brain activity.

**Introduction**

Resting state brain activity is simple to collect, requiring that subjects do nothing in particular for a short period of time (2-7) as they are brain scanned. Resting state brain activity reflects dynamics of the brain (8-10) that are correlated with various kinds of functional networks, including working memory, sensory and motor processes, visual pathways, and many kinds of cognitive/perceptual task related engagement. Often in the context of a specific task (e.g. working memory tasks such as the "n-back" task) the resting state networks (RS-Networks) may fall to a lower level of activity ("deactivation" relative to baseline), although it may also increase in some sub-networks while simultaneously decreasing in others (11). At present, there is no simple phasic/modal or spectral functional account of RS-networks (nonetheless, there has been a considerable amount of modal/clustering and spectral analysis that has been done over the last decade), although a large body of research using preprocessing and systematic controls has minimized the potential claim that RS-networks are a physiologic artifact due to the para-sympathetic nervous system or that they have some simple common cardio-vascular origins. Whatever their specific function may be, they clearly appear to have some fundamental relation to the modulation of the larger scale brain network fluctuations themselves.

It is also known that the RS-networks are not random graphs, but rather appear to be hierarchical in nature and highly structured (12). This structure is often sparse due to the few local "hubs", that are apparent, such that there are a few brain regions with many connections and still a larger number with many fewer connections and so on, and due to the sparse nature are dubbed "small world" networks (13). Some networks that designated "small world" networks are also "scale-free" in that they are also a special type of hub structure that is strongly and quantitatively hierarchical, following a well known probability distribution over hub connections or degree, called the *Pareto* (or Power function distribution).

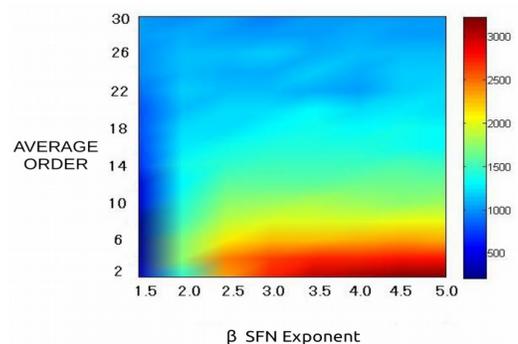

*Figure 1: Adapted from (22) showing the MDS decreasing with Scale Free Exponent, showing a sharp phase transition at 2.0. Implying more fractionated control with exponents >2.0.*

RS-networks are already known to be distinctly different in special populations that demonstrate atypical (AT) brain activity when compared to RS-networks observed in neurotypical (NT) controls. For example, in those individuals with autistic spectrum disorder (ASD) RS-networks have been reported as having "under-connected" (14) long distance connections as well as "over-connected" short distance connections see (15) for discussion of both types). This type of biased connectivity has been associated with reduced "efficiency" in potential communication processes (11). In those individuals with schizophrenia, the RS-networks have been reported, paradoxically, as both over-connected in some sub-networks (parietal-frontal) and under-connected in other sub-networks (parietal-temporal), leading to a potential imbalance in the overall dynamics of their brain networks (11). There are also known qualitative connectivity changes reported in aging, depression, and epilepsy (16, 17, 18). Such network variation is possibly reflected in overall topology and structure of the network and are most likely to be seen in more global measures of network complexity. In fact, Liu et al (19) show systematic changes in degree statistics between typical and schizophrenic networks suggesting that the overall global structure of the hubs in these networks are disrupted in schizophrenia. Whether the network connectivity is atypical by either shifts in under or over connectivity or simultaneously by both kinds of connectivity biases, we are likely to see global shifts in network communication and control. This "communication efficiency" disruption has been formally quantified in the larger field of dynamical network theory based on the sensitivity of Scale-Free behavior. Specifically, it has been shown that as the tail of such distributions become more extreme, such networks tend towards small-world ($\ln(n)$ growth; 20), and as the exponents drop below 2.0 they become "ultra-small world" ($\ln(\ln(n))$ growth), showing highly stable and more efficient communication dynamics.

*Scale-free exponents.* Before we can appreciate the potential neural network mechanisms we are proposing here it is important to first review recent theory in dynamical network complexity. Network complexity/dynamical theory on scale-free networks (21, 22) has shown that there is a dramatic qualitative difference between those scale-free networks with exponents **<2.0** as compared to those **>2.0**. In fact, Nacher and Acuson (22) examine the dynamical control of a network by considering a model of reduced complexity, where a minimum set of possible nodes dominates the whole system-called the minimum dominating set (MDS). The concept of MDS has been applied to the design and/or control of various kinds of discrete systems, which includes mobile ad hoc networks, transportation routing and computer communication networks but not as yet in terms of brain function. It was shown (21) in a series of dynamical network control simulations that the MDS shows a sharp phase transition (see figure 1) at $\beta$=2.0. When $\beta$ < 2.0, the MDS slightly increases or even tends towards constant as the average Degree (the hub connection number) increases. However, the jump in the number of dominating nodes at for $\beta$>2.0 is significantly larger, suggesting a drastic change in the network's dominance phenomena when the scaling exponent begins to approach values near $\beta$ = 2.0   Various network properties when $\beta$<2.0 are consistent with this increase in control, for example the average Degree tends to grow more slowly, (lnln(N)--"ultra small world") with system size (it is not the case that all scale-free networks are small world especially for exponents >3.0). Consequently, with decreasing exponents (<2.0), as in the present case, connections tend to be cheap and hubs tend to be few but in contrast to those networks with exponent (>2.0) considerably more costly to create.   Qualitatively, it has (21) been argued that networks with exponents below 2.0 are similar to an optimized software library reuse system, where modules, which are expensive to create are at the same time very low cost to link and relink depending on specific task requirements.  Consequently, the hubs in SFNs with exponents <2.0 are few, expensive to create and generically productive in scope of use.  Also similarly, it has been (21), shown that in networks, with exponents less than 2.0, contain some type of global control mechanisms that could be used to coordinate global information in the network.  At the minimum this requires some sort of global information exchange mechanisms (from other unmeasured networks), that may not part of the measured network itself but nonetheless allows nodes to interact globally, unlike the case where the exponent >2.0.  In summary, these recent theoretical advances provide explicit predictions that the scale-free exponent at 2.0 is a phase transition, that exponents below 2.0 are in fact "ultra-small world" and this resultant topology predicts decreases in global information and decreased local network control.

*Brain Networks hub structures and SFN exponents:*  The nature of hub structures and their relationship to brain dynamics is also of increasing interest in the neuroimaging research involving resting state brain activity. Many have posited that there are various types of hubs that may be critical in controlling and organizing the emergent brain dynamics.   Bullmore & Sporns (11), for example argue that neuropsychiatric disorders can by thought of as a sort of "dysconnectivity syndrome", depending on the type or nature of the  disruption in topology of the brain networks. They point out that some type of overall connectivity dysfunction should be apparent in the interdependence of the structural and dynamical functional states of those networks.  They further point out that the topology and synchronization as well as most dynamical aspects of functional behavior are strongly affected by the variation in complexity metrics of structural connectivity. If we assume the brain has evolved to maximize communication transfer and minimize wiring costs, then neuropsychiatric disorders that manifest in various ways through connectivity distributional biases, could be identified from the minimum dominating set (MDS), discussed before, through estimates of scale-free exponents. The limitation of the MDS theory is that it can only indicate structural anomalies when the SFN exponents grow strictly greater than 2.0. Nonetheless, this theory as adapted for this particular brain network context could provide diagnostic markers for various neuropsychiatric disorders but not necessarily between disorders, assuming the loss of network communication efficacy is a common property of each disorder. We examine these implications of MDS theory using archival data from Schizophrenics, Autistic Spectrum disorder and compared with two independent samples of neurotypicals.

*Materials and Methods:*

*Resting-state Data*

  *Single individual (RP set).* Repeated scans (23, 24) of a single individual consisted of 66 daily (M, T, TH@ 7:30am) scans contiguously (except for holidays). RS-fMRI was performed in each of the 104 regular scan sessions throughout the data collection period, using a multi-band EPI (MBEPI) sequence lasting 10 minutes (TR=1.16 ms, TE = 30 ms, flip angle = 63 degrees, voxel size = 2.4 mm X 2.4 mm X 2 mm, distance factor=20%, 68 slices, oriented 30 degrees back from AC/PC, 96x96 matrix, 230 mm FOV, MB factor=4).

  *Neurotypical data (BB set)* Neurotypical individuals (25) were scanned while resting state fMRI activity was acquired (The resting-state data were scanned for approximately 8 minutes (500 s) (TR= 2.5 s, TE = 27 ms; acquisition matrix = 64 × 64; flip angle = 77°; slices = 43; voxel size = 3.44 mm × 3.44 mm × 3.40 mm. High resolution MPRAGE anatomical images were also acquired with the scanning parameters as follows: TR = 2530 ms; TE = 3.5 ms; flip angle = 7°; resolution = 1 mm × 1 mm × 1 mm-no gap).

*Schizophrenia data.* We used data from the publicly available COBRE (26) data set. Resting-state scans were 5 minutes in duration (TR= 2000 ms, TE: 29 ms, matrix = 64x64, slices, = 32 voxel size = 3 mm x 3 mm x 4 mm). MPRAGE anatomical images were acquired as well: (TR = 2530 ms; TE=[1.64, 3.5, 5.36, 7.22, 9.08] ms TI =900 ms; flip angle = 7°; resolution =1 mm x 1 mm x 1 mm.

*Autism Spectrum Disorder (ASD set)* We used data collected and archived by the Autism Brian Imaging Exchange (ABIDE) (27) (http://fcon_1000.projects.nitrc.org/indi/abide/) collected at UCLA and randomly sampled subjects from UCLA1 and UCLA2 datasets, resulting in 60 ASD subjects. We age-matched as close as possible to the Biswal set (ABIDE, 8-18, mean=13, SD=3). Resting-state data was collected for 6 minutes (TR = 3000ms; TE = 28 ms, flip angle = 90°; slices = 34; voxel size 3.0 mm x 3.0 mm x 4.0 mm. High resolution MPRAGE images were also acquired: TR=2300 ms; TE = 2.84 ms; flip angle = 9°; resolution = 1 mm x 1 mm x 1.2 mm)

*Preprocessing*

All datasets underwent the same preprocessing steps using FSL which included brain extraction, re-alignment using FSLs MCFLIRT, and registration to the MNI 2mm standard template using FSLs FLIRT (also see Motion section below). ROIs applied to all samples, were developed from a recent analysis of 2200 subjects over 200 tasks (28, 29) which identified common sets of ROIs across a dozen or so different types of tasks. This comprehensive work identified 264 ROIs, subsequently, using similar data sets, other groups (30) have added 19+ more ROIs (giving us 283 total) which we used to extract time series from each data set and are listed in the supplemental materials. The mean time-series for each ROI was used in the subsequent analysis.

*Connectivity.*

There are a number of ways to detect and measure scale-free properties or other network complexity structures. One way is simply to calculate the Pearson "r" correlation between pairs of preprocessed time series, from selected ROIs or voxels. This has been termed "functional connectivity" (7), although it is perhaps better described as simply pairwise covariance, since there is no real connectivity data structure ("skeleton") without norming and thresholding the covariance matrix. Even if the matrix is normed and carefully thresholded, this type of network identification can be unstable and often contain many false positive edges, as shown in simulation studies (31) as well as lacking simple re-sampling validation (32). Moreover, correlations can be produced without any true direct connection between nodes (based on simulations), and so are a poor guide to graph topology. Moreover, correlation estimates are know to be highly sensitive to motion artifacts (32) making their partitioning dependent on the clustering methods used for partitioning (also true for partial correlation).   Consequently, although correlation methods provide some coarse measure of scale-free structure, they generally cannot identify scale-free structure and cannot reliably be used to estimate SF exponents (see supplemental material).

Another common way to estimate networks from time series fluctuations is through conditional independence (33).   Methods of this type can be roughly grouped as "Bayesian network' estimation in that they use patterns of conditional probabilities to infer the underlying network structure.   Recent simulation studies using fMRI-like signals, (34), have shown that Bayes network search algorithms can identify with high probability known networks across at least two degrees of magnitude in network size (5-500 nodes).  One such Bayes network search method, designed specifically for fMRI data, stands out as being particularly effective by using constraints arising from individual differences across sets of brain areas (35) .   IMaGES (Independent Multiple sAmple Greedy Equivalence IMaGES, in fact was benchmarked at 99%/95% , Recall/Precision in adjacency and orientation (as compared with thresholded correlation, 33%, Recall, no orientation; 34).    IMaGES  does a number of pre-processing steps in Degree to increase the time series signal to noise, decrease the drift, and identify significant event change points in each series.   IMaGES is one of the few graph estimation methods that also exploits individual differences as constraint satisfaction opportunities. Thus IMaGES can both estimate common and/or distinctive aspects of ROIs between individual brains.    Because IMaGES is based on conditional probabilities between ROIs across multiple subjects, it is more robust to motion-related artifacts that tend  to induce increases or decreases in correlation in individual subjects often influencing the resulting network structure.  More details concerning IMaGES can be found in a number of papers (35, 36)  that provide more theoretical and algorithmic details.   For each data set, 6 independent samples of 10 subjects  (11 in the RP set) each were prepared and IMaGES was performed separately on each set of 10 subjects resulting in 6 independently estimated graphs.   Subsequently, network complexity using multiple metrics,  including *Efficiency, Diameter, Energy, Global Clustering Coef, Distance Degree Compactness, Graph Index Complexity, Mean Distance Deviation, and Vertex Degree (in/out Degree)* were measured. The **Vertex degree** (the total number of in/out connections of the node) measure in particular, is of interest, since it can be used to assess the hub distribution and  whether the  network displays scale-free behavior.   For each data set, regressions to all 6 samples (of 10 or 11-in the RP case- subjects) were performed where a straight regression line in log-log coordinates  is consistent with a Pareto distribution with negative exponent reflected by the slope of the line.

*Fitting Scale Free Exponents.* Clauset (37) and others have pointed out the potential biases in estimation of power functions using least squares. Simulation studies have shown in very large samples (>10,000) that the class of the function (e.g., power, log-normal, poisson) could be misidentified and the exponent could be biased even if shown to be consistent with the class of power functions. They show that a maximum-likelihood estimator can provide excellent results in recovering known exponents and confirming the power function class using the Komologorov-Smirnov (KS.p) statistic. However, It is less clear as the sample size decreases (where estimates are still desired) whether LSE or MLE are more accurate for exponent estimation (clearly there will be some bias—and standard errors should increase). In fact, earlier work due to in estimation of power (Pareto) functions distributions, due to Richard Quandt (38) showed over a smaller set of samples that there was actually good agreement with sample sizes of 50-2000, between various methods including LSE, MLE and Method of Moments, showing that all estimators are consistent and "practically indistinguishable". Consequently, we report both the least squares estimators and the MLE estimators as well as the power function class probability statistic (KS.p), which for larger values indicates that the distribution is not significantly different from a power function class. Note that in this paper we are primarily interested in reporting the *best power function* exponent fit, in Degree to test the MDS complexity theory discussed above (which for exponents decreasing below 3 or less are predicted to be power functions), despite it being possible to fit other functions given the error variance. In general, we show that the MLE estimator, tends to be more extreme than the LSE estimator with a higher standard error over sets of subjects. We will also use a standard bias correction estimator on the aggregate (6 sets) independent sets later in the paper for yet a 3$^{rd}$ type of scale-free hub estimation. Below in table 1, we report the average value over all six independent samples for all parameters, showing that the power function does, in fact, provide a strong fit to the *Degree* distributions per network data.

*Results*

*Individual repeated scans:* IMaGES was applied to a single subject (RP set; 23, 24) who was scanned for 66 consecutive days. Six samples of 11 subjects each were independently fit with IMaGES after standard preprocessing was applied for drift, motion and intra-session normalization. The unique graph calculated

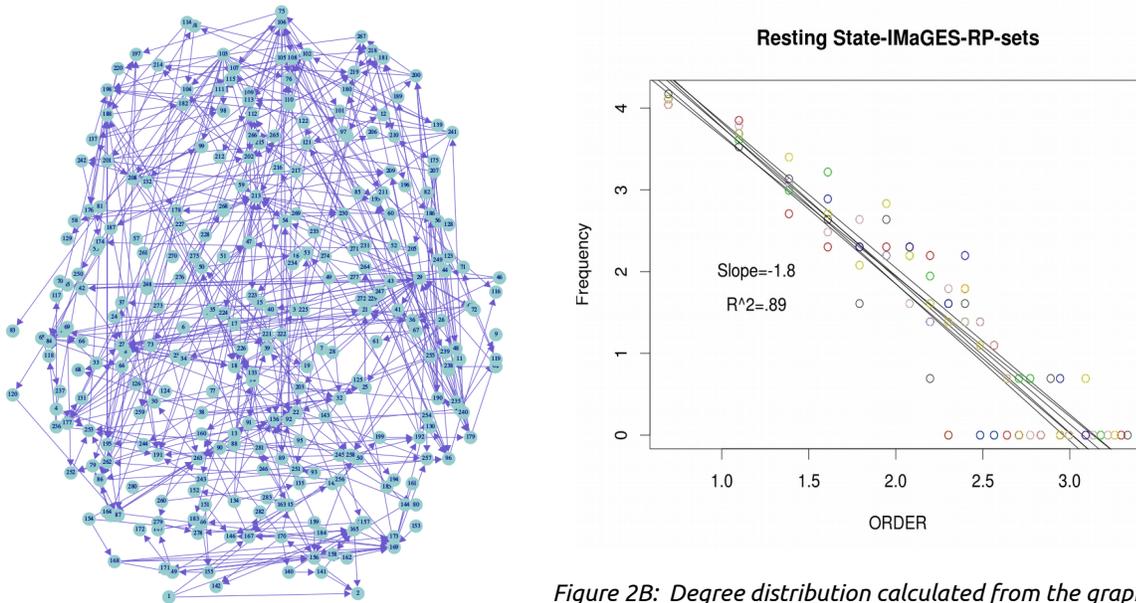

*Figure 2A: Graph structure identified from resting state fMRI data collected over 66 days from a single individual (RP).*

*Figure 2B: Degree distribution calculated from the graph network identified in figure 2A using IMaGES. Each of 6 distribution aggregated over 11 days are fit to a Pareto distribution in log-log coordinates.*

by IMaGES is shown for only one of the six independent samples (11 subjects) in Figure 2A. Examination of the graph in figure 2A, indicates a relatively small number of hubs distributed mainly through temporal, parietal and orbital frontal lobes. Network complexity using multiple indices across all 66 days for each of the six independent sample "individuals" (11 randomly selected per set) was measured. These measures included (we indicate in parentheses the value of one of grouped samples): *Efficiency (0.7), Diameter (8) Energy (349.0) Global Clustering Coef (0.14), Distance Degree Compactness (563574), Graph Index Complexity (0.08), Mean Distance Deviation (103), Vertex order (3) and in/out order.* Each of these measures was estimated over the eleven samples and had remarkable stability, varying over all sets less than 1% over the 66 days. The hub *degree ( not average degree)* measure in particular, is of interest, since it can be used to assess the strength of the network scale-free behavior. The ***degree*** probability distributions (Pareto distribution) of the one graph shown in Figure 2A, is graphed in log-log coordinates in Figure 2B. This figure also shows regressions to all 6 independent samples (different color points) graphed as *degree* distributions with regression in Log-Log coordinates. A straight regression line in these coordinates is consistent with a Pareto distribution with negative exponent reflecting the slope of the line (Note that we also estimated MLE exponents and tested the significance of power function class). The slopes of all distributions are tightly clustered around a mean value of *1.8,* s.e. 0.01, with the total variance accounted for by the regressions for all sets to be about *90%.* The MLE estimates were similar but systematically lower at 1.6, s.e. .02. The

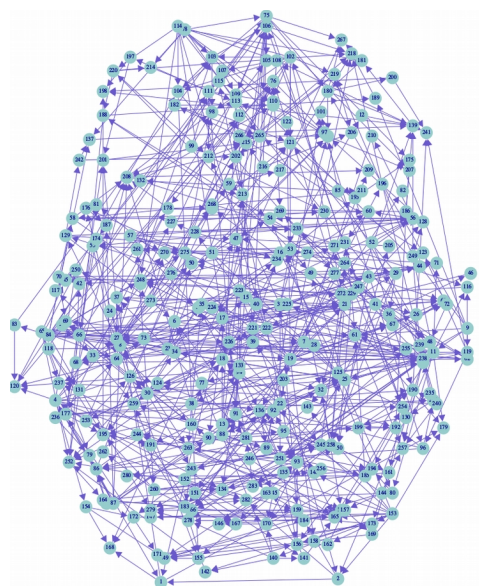 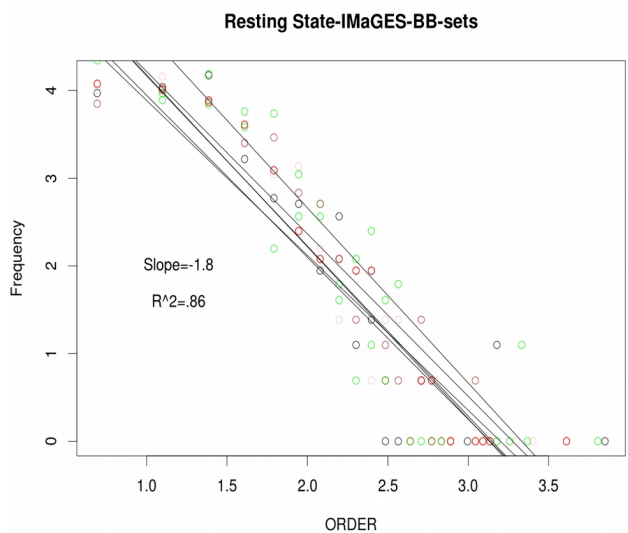

*Figure 3A: Graphical structure identified from Resting state fMRI time series from 60 different individuals. Each distribution is the aggregate of ten individuals into six different samples.*

*Figure 3B: Degree distribution of graphical model identified by IMaGES in figure 3A. Each of 6 distribution s aggregated over 10 individuals each is regressed to a Pareto distribution in log-log coordinates.*

Komologorov-Smirnov "p" (KS.p) statistic was .33 indicating moderate to high support for a power function class. These distributions indicate clearly that the resting state networks are-- given the <2.0 exponents--"ultra-small world" (20). Although networks with exponents less then 2.0 are not uncommon (e.g. protein interaction networks, networks of co-authors etc..) as outlined above they are known to be qualitatively and fundamentally different from those scale-free networks >2.0.

*Between Individual Networks:* In Degree to further establish the statistical validity of the exponent in the brain connectivity, we also randomly sampled 60 neurotypical individuals (BB set) who were (26) scanned while resting state fMRI activity was acquired (The resting-state data were scanned for 500 s with a TR of 2.5 s, resulting in 200 images for each subject. We, again, constructed 6 independent sets of 10 sampled subjects each and submitted each set of 10 subjects independently to IMaGES. From the resultant graphs (see Figure 3A showing one selected graph from the six) complexity analysis and degree was determined for each of the set of 6 graphs and an degree distribution was constructed per set and regressions were found per set all shown in Figure 3B. The resultant complexity measures are similar in value with a few exceptions to the RP repeated scans. For example the following set were estimated, *Efficiency (.8), Diameter (7), Energy (504), Global Clustering Coef (.08), Distance Degree Compactness (481281), Graph Index Complexity (.08), Mean Distance Deviation (85), Vertex order (3.25) and in/out orderagain varying over nodes.* As expected

12the variance across samples is higher then the RP set due to the fact that the sample involves different individuals, particularly in the intercept, but less so in slope. The MLE estimate this time was very similar to the LSE, ( 1.8, .s.e. 0.01). The Degree distribution slope is the same in both individual repeated scans (RP set) and now in the group-wise data (1.8, s.e. 0.01) a Welch two sample *t-test* between the two sets of 6 measures (RP and BB) confirms this (t = -1.5339, df = 10, p-value = 0.1561). Consistent with this slope overlap is the similarity of the various complexity measures including Global clustering coefficient (0.14,0.08), Efficiency (.7,.8) Diameter (7,8), Mean Distance Deviation (85,103 ) and Graph Index complexity (.08,.08). The next set of analyses involve special populations including individuals with Autistic Spectrum Disorder and those with early onset Schizophrenia. Here two questions arise. Given the known network communication and connectivity disruption in special populations, do these networks still show the scale-free signature and does the slope vary within the same range as established before in neurotypicals? We first consider the case of Autistic Spectrum Disorder

*Autistic Spectrum Disorder (ASD)* is a disease that appears to disrupt network communication processes, probably initiating through corpus callosum, and spreading into frontal-parietal networks. In particular, connectivity distributions are shorter than those connections found in neurotypicals, indicating a hyper-connectivity in various regions of the brain (40). We used data collected and archived by the UCLA ABIDE (28) consortium and randomly sampled subjects from Samples 1 and 2, resulting in 60 ASD subjects (*Scanning parameters for ABIDE and COBRE are documented above and also on the database sites*). After fitting the RS-networks with IMaGES we then calculated the standard complexity measures again including: *Efficiency (0.77), Diameter (7), Energy (470), Global Clustering Coef (0.02) Distance Order Compactness (568988), Graph Index Complexity (0.06), Mean Distance Deviation (88), Vertex Order 2.9)* as well as the Degree distribution for each set of 10 sampled individuals, resulting again in 6 independent regressions in Log-Log coordinates shown in figure 4B. On the left of this figure (figure 4A) we show a typical graph derived from 1 set of 10 subjects in the 6 ASD samples. The overall graph structure as identified by IMaGES is similar to the neurotypical graphs in terms of gross features, but as the complexity statistics reveal, there are subtle differences. For example the global clustering coefficient has dropped by a factor of 3 or 4 compared with either neurotypical samples sets, while the index complexity has also decreased slightly. Also the *Degree* distribution exponent has increased from the neurotypicals to 2.4, s.e. .03 also confirmed by the MLE estimate (2.8, s.e .0.1). In general it appears that the overall hub structure for ASD has become more fractionated and is tending towards more localized hubs per region.

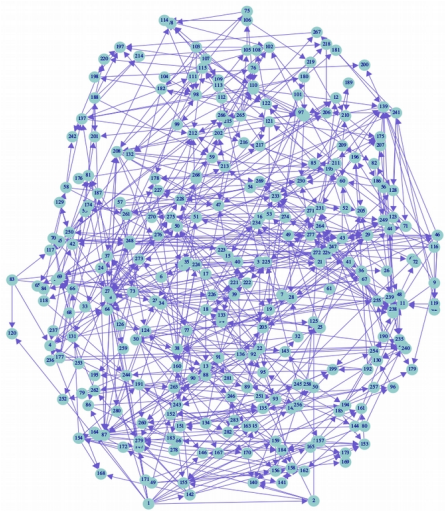

*Figure 4A: Graphical structure identified from 6 different samples of 10 individuals each with Autistic Spectrum Disorder from the UCLA ABIDE database, using samples 1 and 2.*

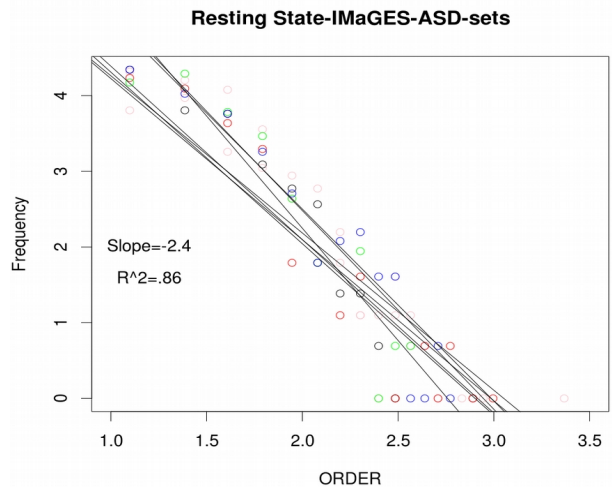

*Figure 4B: Degree distributions from graphical structure in figure 4A, showing 6 different samples aggregated over 10 individuals from the UCLA ABIDE ASD database.*

*Schizophrenic* (SZH). Schizophrenia is yet another devastating mental illness that also has observed shifts in brain connectivity distance distributions. In this case the connectivity distributions show both systematic under-connectivity as well as over-connectivity across various areas of the brain, however with a total global mean shift towards longer distance connections and less hyper-connectivity (41). In contrast to ASD distributions, Schizophrenia tends to have the opposite connectivity distribution, possessing global hypo-connectivity relative to hyper-connectivity in the ASD distance distributions, nonetheless connectivity is disrupted. We assessed the graphical structure using the COBRE (26) data set, randomly sampling 60 subjects from the 80 subjects in the dataset who also were scanned during resting state. These subjects were also again fit with IMaGES over the 6 sets of 10 subjects each, producing an example of one of the resultant graphs shown if Figure 5A. Again we calculated the various complexity measures and Degree distribution from this graph. The complexity measures were not unlike the ASD profile including Efficiency (0.79), Diameter (7), Energy (541), Global Clustering Coef (0.057), Distance Degree Compactness (554156), Graph Index Complexity (0.07), Mean Distance Deviation (73), Vertex Order (3.0) The Order distribution for each of the 6 regressions of each graph is shown in Figure 5B, and shows an average slope value of 2.3, with s.e. 0.04, and a slightly higher MLE exponent of 2.9 with s.e. 0.2. These are values statistically indistinguishable from the ASD scale free distribution exponent of 2.4 ( t = 1.14, df = 9.539, p-value = 0.2805).

The two neurotypical exponents (RP and BB) are significantly different from the two special populations (ASD and SZH) samples (Welch's two-sample t-test:  t = 7.1095, df = 13.119, p-value = 7.549e-06), indicating that the special populations possess scale free distribution exponents greater than 2.0 , with a more than a 30% increase over the neurotypical exponents putting those individuals with atypical brain activity in the scale-free exponent regime of greater than 2.0 (21).   We present all  estimated exponents and related parameters in the next table over all 6 independent sets of 10  (11; RP) subjects each with per cent variance accounted for ($r^2$) and standard errors (se).

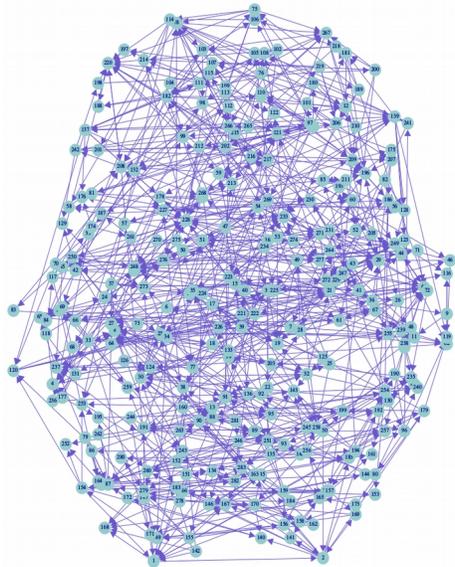

Figure 5A: Graphical structure identified from fMRI time series using IMaGES from 6 samples (one sample shown) aggregated over 10 individuals from the COBRE database.

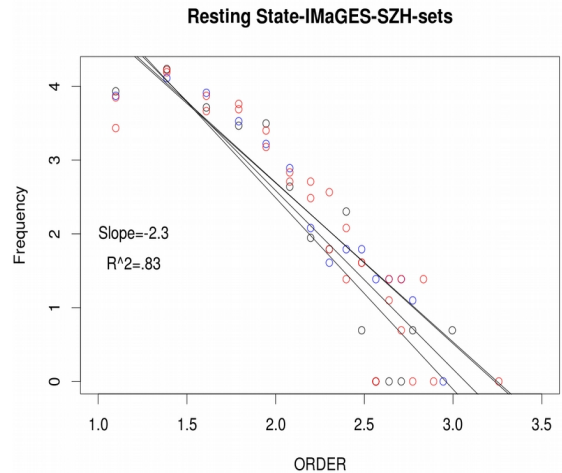

Figure 5B: Degree distributions identified from graphical structure shown in Figure 5A. Each of six samples containing 10 individuals from the COBRE database were regressed against a Pareto distribution in log-log coordinates.

| Parameter Estimates | LSE/se | PVAC/se | MLE/se | KS.p>/se |
|---|---|---|---|---|
| RP | 1.8/.01 | .89/.01 | 1.6/.02 | .33/.09 |
| BB | 1.8/.02 | .84/.02 | 1.8/.02 | .42/.1 |
| ASD | 2.4/.03 | .84/.02 | 2.8/.1 | .77/.06 |
| SZH | 2.3/.04 | .87/.01 | 2.9/.2 | .72/.1 |

Table 1:  Scale Free exponent estimation (LSE and MLE) and standard errors

*Motion.* A possible alternative explanation of the exponent variation might be thought to be in the potential increase in motion artifacts that could create a confound, especially in  atypical populations (33,41). In Degree to remove motion artifacts as a potential explanatory variable, we provide three different kinds of controls and analysis.   First,  in Degree to establish a baseline motion  we  measured the displacement of brain volumes over time using *fsl_motion_outliers,* showing that there was an average displacement of the atypical populations greater than the typical populations (t=3.5,df =12.37, p<0.01).   To test whether this difference could be the source of the SFN exponent difference we observed that there were also significant individual differences in displacement motion, therefore allowing us to sub-sample from the atypical populations  a specific set of *low-motion* subjects  that overlapped with the average displacement distributions for neurotypical populations.   Specifically, we used standard methods for detecting outlier motion in the BOLD time series (using both *fsl_motion_outliers* and *CompCOR; removing "spikes" >1/2 mm*) and selected  sets of 10 subjects each from the ASD sample and the SZH sample with the smallest motion displacement averages that were within the range of the neurotypical subject motion (t=1.2,df=12.5, p>0.15; maximum % average frame displacement >.5mm, NT, 4%, SZH, 5%, ASD, 8%).  In those cases the exponents refit per group and still showed values consistent with each group:   where the exponents were >2.0 (ASD, 2.59, SZH 2.8) despite the absence of significant motion displacement differences between the typical and atypical groups.    This in itself indicates motion has minimal or in fact, no effects on complexity of the hub structure.

Second, in order to further control for  the possibility that motion is a potential  confound, we selected a sample of 10 subjects in the neuro-typical with *high displacement motion* over time and then refit the Pareto and estimated the exponent in this high motion sample . If motion is a factor then we might expect a shift in exponent to be higher in this case, in fact, the exponent was identical  to the original overall motion sample where exponents  were less then 2.0 (i.e., Hi Motion NT, 1.85).   Consequently these tests show that although our atypical samples, do possess more average displacement motion then typical samples, they do not at the same time affect the scale-free exponents, and therefore are *shown to be  mainly independent* of the scale free distributions and their global topology.

Finally, we  more generally tested  the effects of head motion, by simulation tests which were designed with injected noise into the original ASD or SZH time-series. Once adding the noise, we then refit the  distributions and  estimated exponents showing *no change  in the exponents as noise increased up to10% of the mean value of the original time-series.* In these simulations that as long as noise is injected uniformly throughout the network structure, as compared to some systematic noise injected in specific parts of the network structure.  might reduce connectivity globally as overall edge correlation decreases randomly, thus potentially reducing uniformly the number of hubs throughout the topology as opposed to differentially altering the overall topology at either the high or the low end of these distributions, which would have to occur in order to change a topological feature of the graph such as ***degree*** statistics.    So although motion might  be the source of many kinds of first order statistical artifacts (especially within in correlation measures, see 33) and false alarms, it appears not to be a factor in the more complex hub degree statistics estimated using conditional independence methods such as in Bayesian search methods like IMaGES.

*Aggregate Estimates of exponents: controlling for saturation and cutoff bias.*  Finally, we provide a 3$^{rd}$ independent way to estimate the exponents to provide further convergent evidence for the MDS theory.    In this case, we use an estimator often used for fitting SFNs, that can deal with two common biases in the scale-free exponent; one is referred to as *low degree saturation* at the initial part of the distribution which indicates potential mis-estimation or sampling error. The second bias is at the back end of the tail and usually referred to as *high degree cutoff*. Neither bias actually rules out Scale-Free network behavior (1) and leads to a specific SFN distribution estimator that is typically used to detect how the strength of the log-log linearity while controlling for both biases:

$$P(k) \sim (k+k0)^{-\alpha} * (\exp(-k/k1))$$

$$\ln P(k) \sim -\alpha \ln(k+k0) - (k/k1) \quad \text{(linear version)}$$

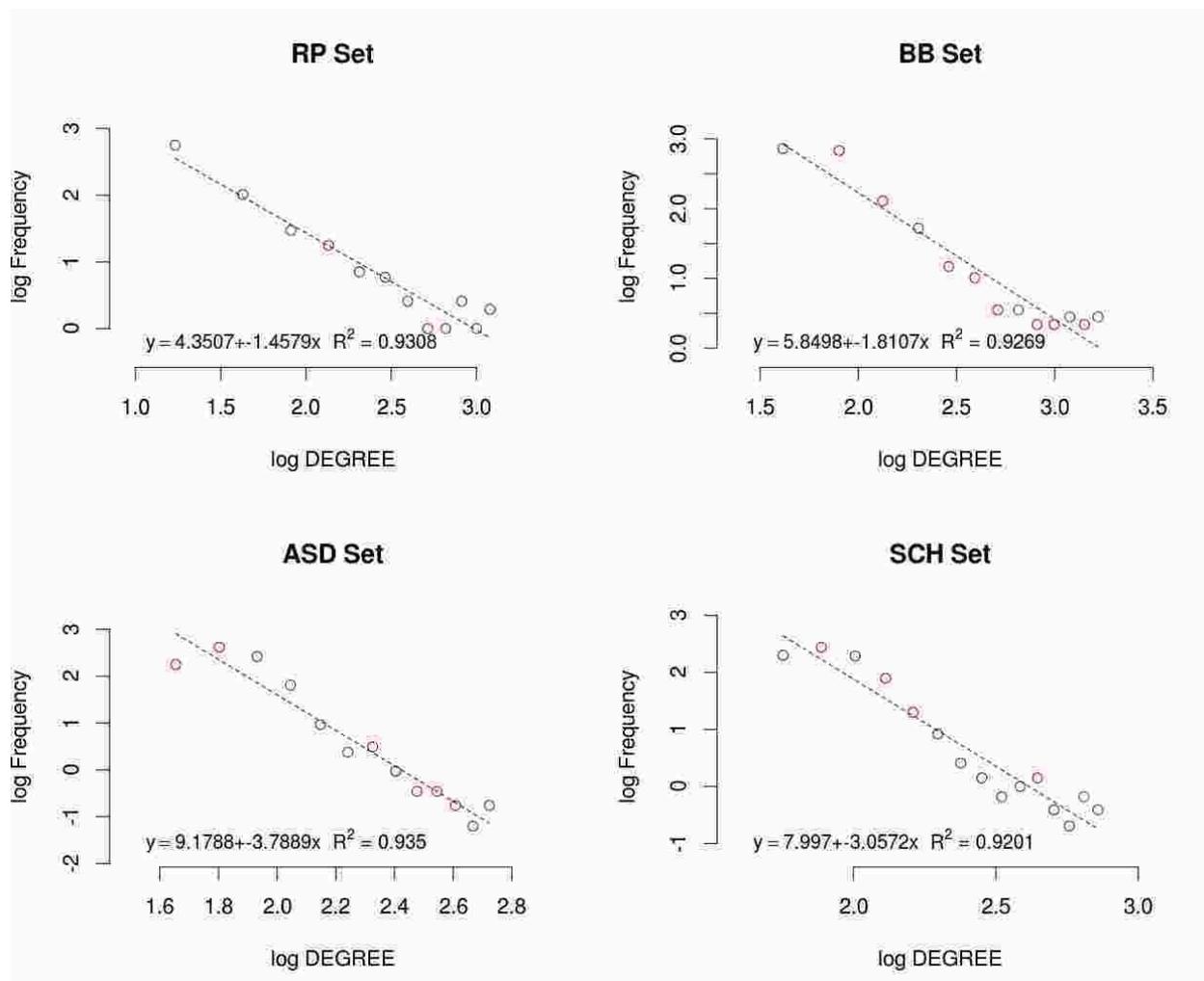

*Figure 6: Aggregate degree distribution over the 6 independent estimates.  Fits are based on "scale-free" estimator with two kinds of bias.  One bias indicated by parameter k0, is termed the high degree saturation bias, while the second bias represented by k1, is called the low degree cut-off bias.  All parameters were fit with nonlinear least squares (NLS) fitter allow all parameters to vary freely.   All fits were excellent with R^2>.92,  Atypical populations showed even greater divergence from 2.0 while neurotypical groups showed values below 2.0 consistent with the MDS theory.*

In this equation, k is the hub degree and the k0, estimates the amount of *low degree saturation* bias, while the k1 term estimates the amount of bias due to *high degree cutoff*. We aggregated the 6 independent samples into one distribution per group and re-estimated the exponents one more time. Shown in Figure 6 are the distributions with the bias correction per group. The bias parameters were left to vary so that the overall bias of each of the log-log fits could be estimated. In the Neurotypical cases, RP and BB both were best fit with with small k0 (0.1, 1.4)) and larger k1(1,1.7) with exponent values still <2.0 **(1.4, 1.8)**, while the Atypical groups were best fit with larger k0 bias terms and smaller k1 (1.9, 0.3; 1.6, 0.5). These estimates were more similar to the MLE differences between the Neurotypical and Atypical groups that crossed 2.0. The Atypical groups, in fact, had considerably higher values **(3.7, 3.0)**, then either MLE or linear LSE estimates but again consistent with the MDS theory that shows phase transition at 2.0.

*Classifier Results.* We also used classifier methods to look for more convergent evidence for the MDS complexity theory. First, we applied a naïve Bayes classifier (nBC) to the exponents from the neurotypical and atypical samples to predict group membership based on the distributions of the exponents in each group.

| Predicted | Actual | |
|---|---|---|
| | NT | AT |
| NT | .92 | .01 |
| AT | .08 | .99 |

Table 2: *Naive Bayes Classifier on exponents*

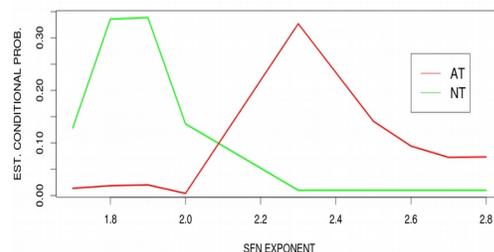

Figure 7: *showing the 10-fold cross validation sampled 100 times. Estimated conditional prob. Dependent on SFN exponent.*

Shown in table 2 is the 10 fold cross-validation showing high probability classification rates for both typical and atypical groups. Using the nBC we can also estimate the conditional probability of each group given the SFN exponent value. In this plot (Figure 7) we can see that the cross-over exponent for equal conditional probability is near 2.1, consistent with the MDS complexity theory outlined before.

*Graph Structure Classification test.* The next classifier test focused on the overall global topology. Can we discriminate the neurotypicals from the atypicals using only the graph structure itself, independent of the SFN exponents? This would provide an independent albeit coarse confirmation (consistency test ) of the scale-free exponent markers in that the overall graph structure itself provides discrimination between neurotypical and atypical groups. To test this possibility we converted each graph to an adjacency vector consisting of of **39,903** (283 x 282)/2 edges indicated by either "0" or "1". We removed

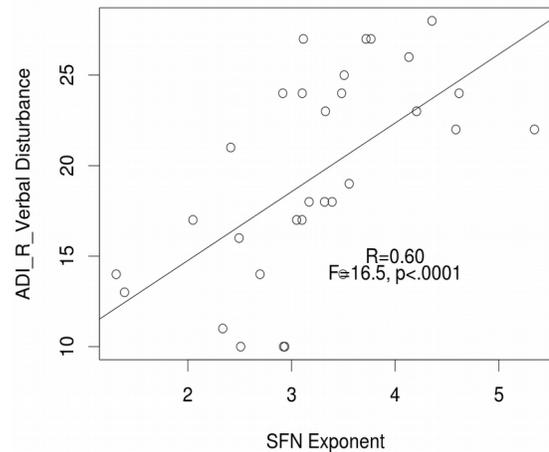

*Figure 8: Verbal Disturbance Measures ("current") as a function of SFN exponent per individual*

nodes that were either all "0" or "1" leaving 15,035 edges to be submitted to SVM (logistic kernel) for both atypical (141) and neurotypical (130) subjects. Using an SVM classifier logistic kernel, the classification overall accuracy based on 10 fold cross-validation was **97%** overall correct accuracy. This result further confirms the graph complexity difference of neurotypical groups from atypical groups based on only the underlying graph structure, thus providing further support that there is a shift in overall topological measures such that SFN exponents might index

*Exponent varies with Phenotypic Severity.* Given the conditional probability dependence on group exponent and individual graph classification results, we further examined whether individual variation in the atypical groups would show dependence on phenotypic measures of the severity of the diseases. Shown in Figure 8 using the ASD sample of the atypical group we created exponents per subject and then correlated them with autistic verbal disturbance measures in particular the ADI_R_Verbal_Total_BV ("current", ABIDE phenotypic measures) showing a strong correlation (R=0.6, p<.0001). No correlation occurred with other measures such as IQ sub-scores or social disturbance measures in the ABIDE phenotypic variables.

Using the COBRE phenotypic measures (there was only one) of diagnostic category (DSM: 295.3-Schizophrenia-paranoid type—both common and at the same time one of the most severe diagnostic categories as compared to Residual type (295.6, for example), we sorted exponents from individuals in the top 10% and bottom 10% and then simply counted the number of Schizophrenia-paranoid per group doing binomial tests showing that the top 10% exponents (>3.0) were significantly dependent on number of Paranoid Schizophrenia in the sample (p<.00001), where the bottom 10% (<3.0) were not (p>.13).

*Discussion.*

We have shown that the identified SFN exponents can potentially provide a biomarker for the atypical brain activity in two special populations. This result depends on a number of innovations and assumptions. First as discussed above, is the recent theoretical work on network complexity showing that network stability and communication are causally related to the hierarchical hub structure. Clearly, stronger modular and cluster structure allows a small group of nodes to provide global communication and synchronization (42). This modularity and communication efficiency is shown by (22) to be indexed by the SFN exponents as they cross a value of **2.0.** A second innovation concerns the specific Bayes network search model—IMaGES—which scales with a large number of nodes (>500), and is sensitive to the conditional dependence and orientation tests. This type of search allows for the node structure to express more subtle hub and modular configurations unlike simpler connectivity measures (e.g correlation) that cannot (see supplemental material on correlations). The orientation graphs are also shown to be robust against head motion that in fact, could be greater in atypical populations. We provide 4 different tests including sub-sampling low motion atypical subjects showing that their SF exponents are unchanged.

Brain mechanisms for this type of scale-free exponent difference between neurotypicals and the two special populations are plausibly related to the types of biases in connectivity distributions that are both apparent, but in different ways in mental diseases such as Schizophrenia and Autistic Spectrum Disorder. Both under-connectivity and over-connectivity can produce shifts in increasing density or sparseness from a normative distribution that typically supports the communication of a larger number of nodes with a smaller control set. Consequently, there might be a number of different ways network communication and control can fail or be disrupted. Of course, we are not therefore arguing that these two diseases are the same, however, that the SFN exponent is unable to distinguish between Autism and Schizophrenic populations, is actually a potentially important outcome. There have been similarities between these two special populations that date back to 1940s, and although the two mental disorders are now considered to be clearly clinically distinct, they also share many clinical features. For example, social withdrawal, communication impairment, and poor eye contact seen in ASD are similar to the negative symptoms seen in youths with Schizophrenia, but again we stress there are many other differences. Nonetheless, If the exponents found for the special populations are clustering the two groups, this would be consistent with a common loss of global network control, and the reduction of potential dynamical control and fractionation of the network hub structures. These shifts in brain network control would also be consistent with disruptions in cognitive control and working memory, hallmarks of the cognitive deficits seen in both of these populations.

*Acknowledgements*

We thank Russell Poldrack (RP data set) for providing his self-scanning data and for general comments on the manuscript. We also thanks Bharat Biswal for providing his resting state data archive (BB data set). We are also grateful to the McDonnell Foundation for its generous support during the development of IMaGES.

*Financial Disclosure*

The Authors have no financial disclosures to make with respect to the submission of this paper.